\documentstyle[12pt]{article}

\begin{document}

\setcounter{page}{1}

\pagestyle{plain} \vspace{1cm}
\begin{center}
\Large{\bf Non-Minimal Inflation Revisited}\\
\small \vspace{1cm} {\bf Kourosh Nozari}\quad and\quad {\bf S.
Shafizadeh}\\
\vspace{0.5cm} {\it Department of Physics, Faculty of Basic
Sciences,\\
University of Mazandaran,\\
P. O. Box 47416-95447, Babolsar, IRAN\\ $^{*}$knozari@umz.ac.ir}
\end{center}
\vspace{1.5cm}
\begin{abstract}
We reconsider an inflationary model that inflaton field is
non-minimally coupled to gravity. We study parameter space of the
model up to the second ( and in some cases third ) order of the
slow-roll parameters. We calculate inflation parameters in both
Jordan and Einstein frames and the results are compared in these two
frames and also with observations. Using the recent observational
data from combined WMAP5+SDSS+SNIa datasets, we study constraint
imposed on our model parameters especially the nonminimal coupling
$\xi$.
\end{abstract}
\newpage
\section{Introduction}
The idea of cosmological inflation is one of the cornerstones of the
modern cosmology. It solves the horizon, flatness and monopole
problems and provides a mechanism for generation of density
perturbations needed to seed the formation of structures in the
universe [1-3]. Inflation is believed to be driven by a scalar field
that can interact essentially with other fields such as gravity. So,
it is natural to include an explicit non-minimal coupling between
inflaton and the gravitational sector. However, incorporation of an
explicit non-minimal coupling has disadvantage that it is harder to
realize inflation even with potentials that are known to be
inflationary in the minimal theory [4]. Using the conformal
equivalence between gravity theories with minimally and
non-minimally coupled scalar fields, for any inflationary model
based on a minimally-coupled scalar field, it is possible to
construct infinitely many conformally related models with a
non-minimal coupling  [5,6] ( see also the more recent reference in
[7]). Non-minimal coupling is forced upon us from several compelling
reasons [4] and a complete theory of cosmological inflation should
take into account possible coupling of the inflaton field and the
gravitational sector of the theory. There are several comprehensive
studies in the field of non-minimal inflation and some severe
constraints are imposed on the values that non-minimal coupling (
especially the conformal coupling) can attain in confrontation with
observational data [4-7]. Nevertheless, the motivation for present
study lies in the fact that there are very limited number of studies
that handled the non-minimal inflation up to the second order of the
slow-roll parameters and even these limited number of studies have
not treated the problem completely ( see for instance [6,7]). On the
other hand, the results of calculations for inflation quantities
preformed in Jordan and Einstein frames are the same just to first
order of the slow-roll parameters [8] and in the second order
calculations there may be considerable differences between results
of the calculations preformed in these two frames. Our purpose here
is to do all of the required calculations up to the second order of
the slow-roll parameters in each frames versus the conformal
coupling. Then by confronting our results with recent observations
from combined WMAP5+SDSS+BAO datasets, we find severe constraint on
the values of the non-minimal coupling $\xi$. We also compare the
results of calculations in this two frames and in this comparison it
is possible to judge about successes of each of these two frames for
explanation of the inflation parameters in comparison with
observations. Our strategy to perform formal and numerical analysis
is to calculate all inflationary parameters up to the second ( in
some cases up to the third) order of the slow-roll parameters versus
the conformal coupling. A numerical analysis of the model parameter
space illuminates the nature of inflationary dynamics of a
non-minimally coupled inflaton field up to second order of the
slow-roll parameters. Finally a detailed discussion on the issue of
frames will be presented.

\section{ Non-Minimal Inflation in the Jordan Frame}
The action of an inflation model where inflaton field is
non-minimally coupled to gravity can be expressed as follows
\begin{eqnarray}
S=\int d^4x \sqrt{-g} \left[\frac{1}{2\kappa^2}R
-\frac{1}{2}g^{\mu\nu}\partial_{\mu}\phi \partial_{\nu}\phi
+\frac{1}{2} \xi R \phi^2+V(\phi)\right].
\end{eqnarray}
Variation of this action with respect to the metric leads to the
Einstein field equations. If we assume a spatially flat FRW
cosmology with line element defined as $ds^{2}=g_{\mu\nu}d x^{\mu}d
x^{\nu}=d t^{2}-a^{2}(t)d \vec{x}^{2}$, the Friedmann equation takes
the following form
\begin{eqnarray}
H^{2}=\frac{\kappa^{2}}{3(1+\kappa^{2}\xi\phi^{2})}\bigg[\frac{1}{2}\dot{\phi}^{2}+V(\phi)-6\xi
H\phi\dot{\phi}\bigg].
\end{eqnarray}
On the other hand, variation of the action with respect to the
scalar field leads to the Klein-Gordon equation of motion [6-8]
\begin{eqnarray}
\ddot{\phi}+3H\dot{\phi}+\bigg[\frac{\kappa^{2}\xi\phi^{2}(1+6\xi)}
{1+\kappa^{2}\xi\phi^{2}(1+6\xi)}\bigg]\frac{\dot{\phi}^{2}}{\phi}
=\frac{1}{1+\kappa^{2}\xi\phi^{2}(1+6\xi)}\bigg[4\kappa^{2}\xi\phi
V(\phi)-(1+\kappa^{2}\xi\phi^{2})V_{,\phi}\bigg].
\end{eqnarray}
We note that existence of a negative conformal coupling ( that is,
$\xi<0$ ) unavoidably introduces two critical values of the scalar
field $\pm \phi_{c}=\pm\frac{1}{\kappa\sqrt{|\xi|}}$,\, which are
barriers that the scalar field cannot cross. Note also that in these
values, the effective gravitational coupling, its gradient, and the
total stress-energy tensor diverge [4,9].

The slow-roll conditions are defined as
\begin{eqnarray}
\bigg|\frac{\ddot{\phi}}{\dot{\phi}}\bigg|\ll H, \quad\quad
\bigg|\frac{\dot{\phi}}{\phi}\bigg|\ll H,\quad\quad
\dot{\phi}^{2}\ll V(\phi), \quad\quad \bigg|\dot{H}\bigg|\ll H^{2}.
\end{eqnarray}
By considering the inflation potential of the type
$V(\phi)=\lambda\phi^{4}$, the slow-roll conditions give
\begin{eqnarray}
3H\dot{\phi}\approx\frac{-4\lambda\phi^{3}}{1+\kappa^{2}\xi\phi^{2}(1+6\xi)},
\end{eqnarray}
and
\begin{eqnarray}
H^{2}\approx\frac{V(\phi)\kappa^{2}}{3(1+\kappa^{2}\xi\phi^{2})}.
\end{eqnarray}
In the Jordan frame, the slow-roll parameters which are defined as
\begin{eqnarray}
\epsilon\equiv\frac{-\dot{H}}{H^{2}},\quad\quad
\eta\equiv\frac{-\ddot{H}}{H\dot{H}},\quad\quad
\zeta\equiv\frac{V'\delta\phi}{\dot{\phi}^{2}}=\frac{V'H}{2\pi\dot{\phi}^{2}}\,\,,
\end{eqnarray}
can be determined as follows
\begin{eqnarray}
\epsilon\simeq\frac{4(2+\kappa^{2}\xi\phi^{2})}
{\kappa^{2}\phi^{2}(1+\kappa^{2}\xi\phi^{2}(1+6\xi))}\,\,,
\end{eqnarray}
\begin{eqnarray}
\eta\simeq\frac{4(2-\kappa^{2}\xi\phi^{2})}{\kappa^{2}\phi^{2}(2+\kappa^{2}\xi\phi^{2})}
\end{eqnarray}
and
\begin{eqnarray}
\zeta\simeq\frac{(3\lambda)^{\frac{1}{2}}}{8\pi\xi^{\frac{3}{2}}}
(1+\kappa^{2}\xi\phi^{2}(1+6\xi))^{2}.
\end{eqnarray}
During the inflationary phase, each of these slow-roll parameters
remains less than unity. Solving for the value of the field at the
time of horizon-crossing is difficult in either frame, but following
[6], we can use the fact that scales of interest to us crossed
outside of the horizon approximately $70$ e-folds before the end of
inflation, that is
\begin{eqnarray}
e^{\alpha}\equiv\frac{a(t_{end})}{a(t_{HC})}\sim e^{60},
\end{eqnarray}
where HC marks horizon crossing.  $\phi_{HC}$ which is needed to
evaluate the spectral index $n_{s}$ can be calculated using equation
(11). To do this end, we should specify $\phi_{end}$, the value of
the Jordan-frame field at the time that inflation ends. As usual,
inflation ends when the condition $\epsilon=1$ is fulfilled (
however, we note that it is possible also to fulfill $\eta=1$ even
before fulfilling the condition $\epsilon=1$. In this case the
condition for exiting from inflation phase is given by $\eta=1$).
Setting $\epsilon=1$, we arrive at the condition $\kappa^{2}\xi
\phi_{end}^{2}=\beta^{2}(\xi)$ where $\beta$ is given as follows
\begin{equation}
\beta=\bigg(\frac{(12\xi-1)+\sqrt{432\xi^{2}+24\xi+1}}{2(1+6\xi)}\bigg)^{\frac{1}{2}}.
\end{equation}
Now the slow-roll parameters can be expressed just versus the
non-minimal coupling as follows
\begin{eqnarray}
\epsilon\simeq\frac{4\xi}{m^{2}}\frac{2+m^{2}}{1+m^{2}(1+6\xi)}\,\,,
\end{eqnarray}
\begin{eqnarray}
\eta\simeq\frac{4\xi}{m^{2}}\frac{2-m^{2}}{2+m^{2}}\,\,,
\end{eqnarray}
and
\begin{eqnarray}
\zeta\simeq\frac{(3\lambda)^{\frac{1}{2}}}{8\pi\xi^
{\frac{3}{2}}}(1+m^{2}(1+6\xi))^{2}\, ,
\end{eqnarray}
where $m^{2}(\xi)=\kappa^{2}\xi \phi_{HC}^{2}$ and the appropriate
$m(\xi)$ can be determined by the initial conditions. Using the
definition of $\alpha$ and up to the first order of the non-minimal
coupling, in the limit of $m\gg 1$ we find
\begin{eqnarray}
\epsilon\simeq\frac{8\xi}{(16\xi\alpha+1)}\,\,,
\end{eqnarray}
\begin{eqnarray}
\eta\simeq\frac{-8\xi}{(3+16\xi\alpha)}\,\,,
\end{eqnarray}
and
\begin{eqnarray}
\zeta\simeq\frac{(3\lambda)^{\frac{1}{2}}}{16\pi\xi^
{\frac{3}{2}}}(16\xi\alpha+1)^{2}\, .
\end{eqnarray}
In this case, the first order ( in the slow-roll parameters)
spectral index, $n_{s}= 1-6\epsilon+2\eta$ can be written as
\begin{eqnarray}
n_{s}=1-\frac{8}{\kappa^{2}\phi^{2}}\bigg(\frac{3(2+\kappa^{2}\xi\phi^{2})}{\big
(1+\kappa^{2}\xi\phi^{2}(1+6\xi)\big)}+\frac{(2-\kappa^{2}\xi\varphi^{2})}{(2+\kappa^{2}\phi^{2})}\bigg)
\end{eqnarray}
\begin{eqnarray}
n_{s}\simeq1-\frac{32\xi}{16\xi\alpha-1}\,\,,
\end{eqnarray}
where we have used the approximation $\alpha\simeq 60 \gg 1$. Up to
the second order of the slow-roll parameters, the spectral index
$n_{s}$ depends on the three parameters $\epsilon$,\, $\eta$\,  and
$\zeta$ in the following manner [6]
\begin{eqnarray}
n_{s}\simeq 1-6\epsilon+2\eta+\frac{1}{3}(44-18c)\epsilon^{2}+
(4c-14)\epsilon\eta+\frac{2}{3}\eta^{2}+\frac{1}{6}(13-3c)\zeta^{2}\,,
\end{eqnarray}
where $c\equiv4(\ln2+\gamma)\simeq 5.081$  with $\gamma\simeq0.577$.
Therefore, the spectral index in the Jordan frame calculated up to
the second order of the slow-roll parameters expressed in terms of
the non-minimal coupling is as follows

$$n_{s}\simeq1-6\Big(\frac{8\xi}{16\xi\alpha+1}\Big)+2\Big(\frac{-16\xi}{3+16\xi\alpha}\Big)+
6.324\Big(\frac{8\xi}{16\xi\alpha+1}\Big)\Big(\frac{-16\xi}{3+16\xi\alpha}\Big)$$
$$- 15.819\Big(\frac{8\xi}{16\xi\alpha+1}\Big)^{2}+
\frac{2}{3}\Big(\frac{-16\xi}{(3+16\xi\alpha)}\Big)^{2}$$
\begin{eqnarray}
+0.373\bigg(\frac{(3\lambda)^{\frac{1}{2}}}{16\pi\xi^
{\frac{3}{2}}}(16\xi\alpha+1)^{2}\bigg)^{2}.
\end{eqnarray}
Since by definition $$\alpha_{s}=\frac{d n_{s}}{d\
lnk}=\frac{d\phi}{d \ ln k}\frac{d n_{s}}{d \phi}$$ and
$$\frac{d\  ln k}{d \phi}=\frac{4\pi}{m_{pl}^{2}}\frac{H}{H'}(\epsilon-1),$$
the running of the spectral index in the Jordan frame can be
calculated using equations (16)-(18) to find
$$\alpha_{s}=14\epsilon\eta-12\epsilon^{2}-2\zeta^{2}-2\Big(-18c+\frac{151}{3}\Big)\epsilon^{2}\eta
-2\Big(\frac{-44}{3}+4c\Big)\epsilon\eta^{2}$$
\begin{eqnarray}
-5\Big(c-\frac{11}{3}\Big)\epsilon\zeta^{2}-\frac{1}{2}\Big(7-c\Big)\eta\zeta^{2}
-\frac{4}{3}\Big(44-18c\Big)\epsilon^{3}.
\end{eqnarray}
The amplitude of the scalar perturbation is defined as
\begin{eqnarray}
P_{s}(k)=\frac{k^{3}}{2\pi^{2}}\frac{H^{2}}{\dot{\phi}^{2}}\big|\delta\phi_{k}\big|^{2}
\equiv A_{s}^{2}\bigg(\frac{k}{aH}\bigg)^{n_{s}-1},
\end{eqnarray}
where
\begin{eqnarray}
A_{s}^{2}=\frac{1}{2\pi^{2}}\bigg(\frac{H^{2}}{\dot{\phi}}\bigg)^{2}.
\end{eqnarray}
The amplitude of the tensor perturbations at the Hubble crossing are
given by
\begin{eqnarray}
A_{T}^{2}=\frac{8}{m_{pl}^{2}}\bigg(\frac{H}{2\pi}\bigg)^{2}.
\end{eqnarray}
Therefore, the tensor-to-scalar ratio in our setup is given by
\begin{eqnarray}
r=16\frac{A_{T}^{2}}{A_{s}^{2}}=\frac{128}
{\kappa^{2}\phi^{2}}\frac{(1+\kappa^{2}\xi\phi^{2})^{2}}{\big(1+\kappa^{2}\xi\phi^{2}(1+6\xi)\big)^{2}}\,\,,
\end{eqnarray}
where transform to the following form
\begin{eqnarray}
r\simeq\frac{256\xi}{(16\xi\alpha-1)(1+6\xi)}.
\end{eqnarray}
Finally, the number of e-folds in Jordan frame is defined as follows
\begin{eqnarray}
N=\int_{\phi_{i}}^{\phi_{end}}\frac{H}{\dot{\phi}}d\phi=
-\frac{1}{4}\int_{\phi_{i}}^{\phi_{end}}\frac{\phi[1+\kappa^{2}\xi\phi^{2}(1+6\xi)]}
{1+\kappa^{2}\xi\phi^{2}}d\phi.
\end{eqnarray}
Since $\kappa^{2}\xi \phi^{2}\gg 1$, this integral can be calculated
to find
$$ N\approx \frac{\kappa^{2}(1+6\xi)}{8}
(\phi_{i}^{2}-\phi_{end}^{2}).$$  In a viable inflation scenario, we
can neglect $\phi_{end}^{2}$ in comparison with $\phi_{i}^{2}$. So
we find
$$N\approx \frac{\kappa^{2}(1+6\xi)}{8} \phi_{i}^{2}.$$ Most of the
inflationary models give $N\geq60$. With this number of e-folds, one
can find an estimation of the value of $\phi_{i}$ if the value of
$\xi$ is given. If we consider $0.0003\leq \xi \leq 0.166$ ( will be
discussed later), we find $\phi_{i}\geq 22 M_{pl}$ in our setup.

We note that for a minimally coupled $(\xi=0)$ spectral index and
the tensor-to-scaler ratio are calculated in the standard manner:
\begin{eqnarray}
n_{s}=1-\frac{2}{\alpha}\,\,\,\,,\,\,\,\,r=1-\frac{16}{\alpha}\,\,\,\,.
\end{eqnarray}
In the next section we repeat all of the above calculations now in
the Einstein frame.

\section{ The Analysis in the Einstein Frame}
Now we reconsider the previous analysis but now in the Einstein
frame. We adopt the following conformal transformation
\begin{eqnarray}
\hat{g}_{\mu\nu}=\Omega g_{\mu\nu}\,\,\,\,\,\,,\,\,\,\,\,\,\Omega
\equiv 1+\kappa^{2}\xi\phi^{2},
\end{eqnarray}
where we use a hat on variables defined in the Einstein frame. With
this conformal transformation, we find
\begin{eqnarray}
S=\int d^{4}x\sqrt{-\hat{g}}\
\bigg[\frac{1}{2\kappa^{2}}\hat{R}-\frac{1}{2}F^{2}(\phi)
\hat{g}^{\mu\nu}\partial_{\mu}\phi\partial_{\nu}\phi+\hat{V}(\phi)\bigg],
\end{eqnarray}
where by definition
\begin{eqnarray}
F^{2}(\phi)\equiv\frac{1+\kappa^{2}\xi\phi^{2}(1+6\xi)}{(1+k^{2}\xi\phi^{2})^{2}}
\end{eqnarray}
and
\begin{eqnarray}
\hat{V}(\hat{\phi})\equiv\frac{V(\phi)}{(1+\kappa^{2}\xi\phi^{2})^{2}}\,
.
\end{eqnarray}
We transform our coordinate system so that
\begin{eqnarray}
\hat{a}=\sqrt{\Omega}a\,\,\,\,\,\,\,\,\,,\,\,\,\,d\hat{t}=\sqrt{\Omega}dt,
\end{eqnarray}
and in this case the metric takes the usual
Friedmann-Robertson-Walker form
\begin{eqnarray}
d\hat{s}^{2}=d\hat{t}^{2}-\hat{a}^{2}(\hat{t})d \vec{x}^{2}.
\end{eqnarray}
The physical quantities in the Einstein frame should be defined in
this coordinate system. The field equations in this frame are given
as follows
\begin{eqnarray}
\hat{H}^{2}=\frac{\kappa^{2}}{3}\bigg[\Big(\frac{d\hat{\phi}}{d\hat{t}}\Big)^{2}+\hat{V}(\hat{\phi})\bigg],\quad\quad\quad
\frac{d^{2}\hat{\phi}}{d\hat{t}^{2}}+3\hat{H}\frac{d\hat{\phi}}{d\hat{t}}+\frac{d\hat{V}}{d\hat{\phi}}=0,
\end{eqnarray}
where
\begin{eqnarray}
\hat{H}\equiv\frac{1}{\hat{a}}\frac{d\hat{a}}{d\hat{t}}\frac{1}{\sqrt{\Omega}}
\Big(H+\frac{1}{2}\frac{\dot{\Omega}}{\Omega}\Big), \quad\quad
\frac{d\hat{\phi}}{d\hat{t}}\equiv\Big(\frac{d\hat{\phi}}{d
\phi}\Big)\Big(\frac{d
t}{d\hat{t}}\Big)\dot{\phi}=\frac{\sqrt{1+\kappa^{2}\xi\phi^{2}(1+6\xi)}}{\Omega^{\frac{3}{2}}}\dot{\phi}\,\,.
\end{eqnarray}
Under the slow-roll approximations  $\dot{\hat{\phi}}^{2}\ll
\hat{V}$ and $\ddot{\hat{\phi}}\ll3\hat{H}\dot{\hat{\phi}}$, we find
\begin{eqnarray}
\hat{H}^{2}=\frac{\kappa^{2}}{3}\hat{V}(\hat{\phi}), \quad\quad\quad
3\hat{H}\frac{d\hat{\phi}}{d\hat{t}}+\frac{d\hat{V}}{d\hat{\phi}}=0,
\end{eqnarray}
respectively. We define the slow-roll parameters in the Einstein
frame as
$\hat{\epsilon}\equiv\frac{1}{2\kappa^{2}}\Big(\frac{\hat{V}^{\prime}(\hat{\phi})}{\hat{V}(\hat{\phi})}\Big)^{2}$,\,
$\hat{\eta}\equiv\frac{1}{\kappa^{2}}\Big(\frac{{\hat{V}}''(\hat{\phi})}{\hat{V}(\hat{\phi})}\Big)$\,
and $\hat{\zeta}\equiv\frac{1}{\kappa^{2}}
\Big(\frac{\hat{V}^{\prime}(\hat{\phi}){\hat{V}}''(\phi)}{\hat{V}(\hat{\phi})}\Big)$,
where a prime marks the differentiation with respect to
$\hat{\phi}$. For  $V(\phi)=\lambda\phi^{4}$\, as the inflaton
potential and suitable transformations as mentioned above, we find (
see also [6])
\begin{eqnarray}
\hat{\epsilon}=\frac{8}{\kappa^{2}\phi^{2}\Big(1+\kappa^{2}\xi\phi^{2}(1+6\xi)\Big)},
\end{eqnarray}
and
\begin{eqnarray}
\hat{\eta}=\frac{12\phi^{-2}(1+\kappa^{2}\xi\phi^{2})}{\kappa^{2}\Big(1+\kappa^{2}\xi\phi^{2}(1+6\xi)\Big)}
-\frac{4\kappa^{2}\xi(1+6\xi)(1+\kappa^{2}\xi\phi^{2})}{\Big(1+\kappa^{2}\xi\phi^{2}(1+6\xi)\Big)^{2}}-
\frac{16\kappa^{2}\xi}{\Big(1+\kappa^{2}\xi\phi^{2}(1+6\xi)\Big)}.
\end{eqnarray}
If we write $\kappa^{2}\xi\phi_{HC}^{2}=m^{2}(\xi)$, then the
slow-roll parameters can be rewritten as follows
\begin{eqnarray}
\hat{\epsilon}=\frac{8\xi}{m^{2}(1+m^{2}(1+6\xi))},
\end{eqnarray}
\begin{eqnarray}
\hat{\eta}=4\xi\frac{3+m^{2}(1+12\xi)-2m^{4}(1+6\xi)}{m^{2}(1+m^{2}(1+6\xi))^{2}},
\end{eqnarray}
and
\begin{eqnarray}
\hat{\zeta}=4
\sqrt{2}\xi\frac{|3+2m^{2}(-2+3\xi)-15m^{4}(1+6\xi)-6m^{6}(1+6\xi)^{2}+2m^{8}(1+6\xi)^{2}|^{\frac{1}{2}}}
{m^{2}(1+m^{2}(1+6\xi))^{2}}
\end{eqnarray}
respectively. These equations can be rewritten in terms of $\xi$ and
$\alpha$ as follows
\begin{eqnarray}
\hat{\epsilon}=\frac{32\xi}{(16\xi\alpha-1)(16\xi\alpha+1)}\,,
\end{eqnarray}
\begin{eqnarray}
\hat{\eta}\simeq\frac{-16\xi}{16\xi\alpha-1}\,,
\end{eqnarray}
\begin{eqnarray}
\hat{\zeta}\simeq\frac{16\xi}{16\xi\alpha-1}
\end{eqnarray}
respectively. The first order result,
$\hat{n}_{s}=1-6\hat{\epsilon}+2\hat{\eta}$, can be approximated in
the limit of $m\gg1$ to find
\begin{eqnarray}
\hat{n_{s}}\simeq1-\frac{16\xi}{m^{2}(1+6\xi)},
\end{eqnarray}
which can be rewritten as follows
\begin{eqnarray}
\hat{n_{s}}\simeq1-\frac{32\xi}{16\xi\alpha-1}.
\end{eqnarray}
The spectral index to the second order in this frame can be obtained
as follows
$$\hat{n}_{s}\simeq1-\frac{48\xi}{m^{4}(1+6\xi)}-\frac{16\xi}{m^{2}
(1+6\xi)}-15.819\frac{64\xi^{2}}{m^{8}(1+6\xi)^{2}}
+6.324\frac{-64\xi^{2}}{m^{6}(1+6\xi)^{2}}$$
\begin{eqnarray}
+\frac{128}{3}\frac{\xi^{2}}{m^{4}(1+6\xi)^{2}}+
0.373\frac{64\xi^{2}}{m^{4}(1+6\xi)^{2}}
\end{eqnarray}
On the other hand, the second order result for $n_{s}$ can be
approximated in the limit of $m\gg 1$ to find
$$\hat{n}_{s}\simeq1-6\Big(\frac{32\xi}{(16\xi\alpha-1)(16\xi\alpha+1)}\Big)+
2\Big(\frac{-16\xi}{16\xi\alpha-1}\Big)+
6.324\Big(\frac{32\xi}{(16\xi\alpha-1)(16\xi\alpha+1)}\Big)\Big(\frac{-16\xi}{-1+16\xi\alpha}\Big)$$
$$- 15.819\Big(\frac{32\xi}{(16\xi\alpha-1)(16\xi\alpha+1)}\Big)^{2}+
\frac{2}{3}\Big(\frac{-16\xi}{(-1+16\xi\alpha)}\Big)^{2}$$
\begin{eqnarray}
+0.373\bigg(\frac{16\xi}{16\xi\alpha-1}\bigg)^{2}\,\,.
\end{eqnarray}
The running of the spectral index in the Einstein frame in our setup
is given by
$$\hat{\alpha}_{s}=14\hat{\epsilon}\hat{\eta}-12\hat{\epsilon}^{2}
-2\hat{\zeta}^{2}-2(-18c+\frac{151}{3})\hat{\epsilon}^{2}\hat{\eta}
-2(\frac{-44}{3}+4c)\hat{\epsilon}\hat{\eta}^{2}$$
\begin{eqnarray}
-5(c-\frac{11}{3})\hat{\epsilon}\hat{\zeta}^{2}-\frac{1}{2}(7-c)\hat{\eta}\hat{\zeta}^{2}
-\frac{4}{3}(44-18c)\hat{\epsilon}^{3}.
\end{eqnarray}
This relation can be translated to the following expression
$$\hat{\alpha}_{s}=14\bigg(\frac{32\xi}{(16\xi\alpha-1)(16\xi\alpha+1)}\bigg)
(\frac{-16\xi}{16\xi\alpha-1})-12\bigg(\frac{32\xi}{(16\xi\alpha-1)(16\xi\alpha+1)}
\bigg)^{2}-2(\frac{16\xi}{16\xi\alpha-1})^{2}$$
$$-2(-18c+\frac{151}{3})\bigg(\frac{32\xi}{(16\xi\alpha-1)(16\xi\alpha+1)}
\bigg)^{2}(\frac{-16\xi}{16\xi\alpha-1})
-2(\frac{-44}{3}+4c)\bigg(\frac{32\xi}{(16\xi\alpha-1)(16\xi\alpha+1)}\bigg)$$
$$(\frac{-16\xi}{16\xi\alpha-1})^{2}-5(c-\frac{11}{3})\bigg(\frac{32\xi}{(16\xi\alpha-1)
(16\xi\alpha+1)}\bigg)(\frac{16\xi}{16\xi\alpha-1})^{2}
-\frac{1}{2}(7-c)(\frac{-16\xi}{16\xi\alpha-1})(\frac{16\xi}{16\xi\alpha-1})^{2}$$
\begin{eqnarray}
-\frac{4}{3}(44-18c)\bigg(\frac{32\xi}{(16\xi\alpha-1)(16\xi\alpha+1)}\bigg)^{3}.
\end{eqnarray}
The tensor-to-scalar ratio in the Einstein frame is given as follows
\begin{eqnarray}
\hat{r}=16\frac{\hat{A_{T}^{2}}}{\hat{A_{R}^{2}}}\simeq\frac{512\xi}{(16\xi\alpha-1)(16\xi\alpha+1)}\,\,.
\end{eqnarray}
Finally, the number of e-folds in the Einstein frame is given by
\begin{eqnarray}
\hat{N}=\int_{\hat{\phi}_{i}}^{\hat{\phi}_{end}}\bigg(\frac{\hat{H}}{\dot{\hat{\phi}}}\bigg)d\hat{\phi}.
\end{eqnarray}
After construction of the mathematical framework, in the next
section we study outcomes of this setup in a numerical procedure.

\section{Numerical Analysis: Comparison with Observations}
Table $1$ gives a comparison between our numerical results in two
frames and also in two different order of approximations. The
combined WMAP5+SDSS+BAO maximum likelihood (ML) value of $n_{s}$ is
given by $n_{s}^{(ML)}=0.962$, while the combined WMAP5+SDSS+BAO
mean value (MV) is given by $n_{s}^{(MV)}=0.960\pm0.013$\, [10].
Obviously, the second order results have better agreement with ML
result and all results lie in the range of the mean values. We note
that in the construction of this table we have used the COBE
normalization to find the appropriate value of $\lambda$ so that
\begin{equation}
\bigg(\frac{\lambda}{\xi^{2}}\bigg)^{\frac{1}{2}}\approx
\frac{1}{2}\bigg(\frac{\delta\rho}{\rho}\bigg)_{hor}=\bigg(\frac{\delta
T}{T}\bigg)_{rms}=1.1\times10^{-5}
\end{equation}
which by adopting the conformal coupling, $\xi=\frac{1}{6}$, gives
$\lambda\approx3.4\times 10^{-12}$. Table $2$ compares the values of
the running of the spectral index in two different frames and also
in two different order of approximations. Our analysis shows that
this non-minimal model generally gives negative values of the
running of the spectral index. Also, up to the second order of the
slow-roll parameters, the values of the running of the spectral
index in the Jordan frame are larger than the values of
corresponding quantity in the Einstein frame. Table $3$ compares the
values of the tensor-to-scalar ratio in two frames and also in
different order of approximations. The value of $r$ in the Jordan
frame is more close to observational result than the corresponding
result in the Einstein frame.
\begin{table}[htp]
\begin{center}
\caption{Constraining the non-minimal inflation with WMAP5: The
Spectral Index} \vspace{0.5 cm}
\begin{tabular}{|c|c|c|c|c|c|c|c|c|c|c|}

\hline
\hline $$&$n_s(First\,\,order)$&$n_{s}(Second\,\, order)$\\
\hline $Einstein \,\,Frame$ &$0.966$&
$0.964$\\
\hline $Jordan \,\,Frame$& $0.966$
& $0.960$ \\
\hline $WMAP5+SDSS+BAO \,\,Mean\,\, Value$&$0.960\pm0.013$&$---$\\
\hline $WMAP5+SDSS+BAO \,\,\,ML$&$0.962$&$---$\\
 \hline
\end{tabular}
\end{center}
\end{table}

\begin{table}[htp]
\begin{center}
\caption{Constraining the non-minimal inflation with WMAP5: The
Running of the Spectral Index} \vspace{0.5 cm}
\begin{tabular}{|c|c|c|c|c|c|c|c|c|c|c|}

\hline
\hline $$&$\alpha_s(First\,\,order)$&$\alpha_{s}(Second\,\, order)$\\
\hline $Einstein \,\,Frame$ &$-1.7\times10^{-3}$&
$-0.015$\\
\hline $Jordan \,\,Frame$& $-1.7\times10^{-3}$
& $-0.013$ \\
\hline $WMAP5+SDSS+BAO\,\, MV$&$-0.068<dn_{s}/d\ln k<0.012$&$---$\\
 \hline
\end{tabular}
\end{center}
\end{table}

\begin{table}[htp]
\begin{center}
\caption{Constraining the non-minimal inflation with WMAP5:
Tensor-to-Scalar Ratio} \vspace{0.5 cm}
\begin{tabular}{|c|c|c|c|c|c|c|c|c|c|c|}

\hline \hline $$&$r$\\
 \hline $Einstein \,\,Frame$
&$0.003$\\

\hline $Jordan \,\,Frame$& $0.134$\\

\hline $WMAP5+SDSS+BAO\,\, MV$&$r<0.22$\\
 \hline
\end{tabular}
\end{center}
\end{table}

Figure $1$ shows the variation of the slow-roll parameter $\epsilon$
versus $\xi$ in two frames. Obviously, in the presence of the
non-minimal coupling, natural exit from the inflation phase is
achieved without any additional mechanism.
\begin{figure}[htp]
\begin{center}
\includegraphics{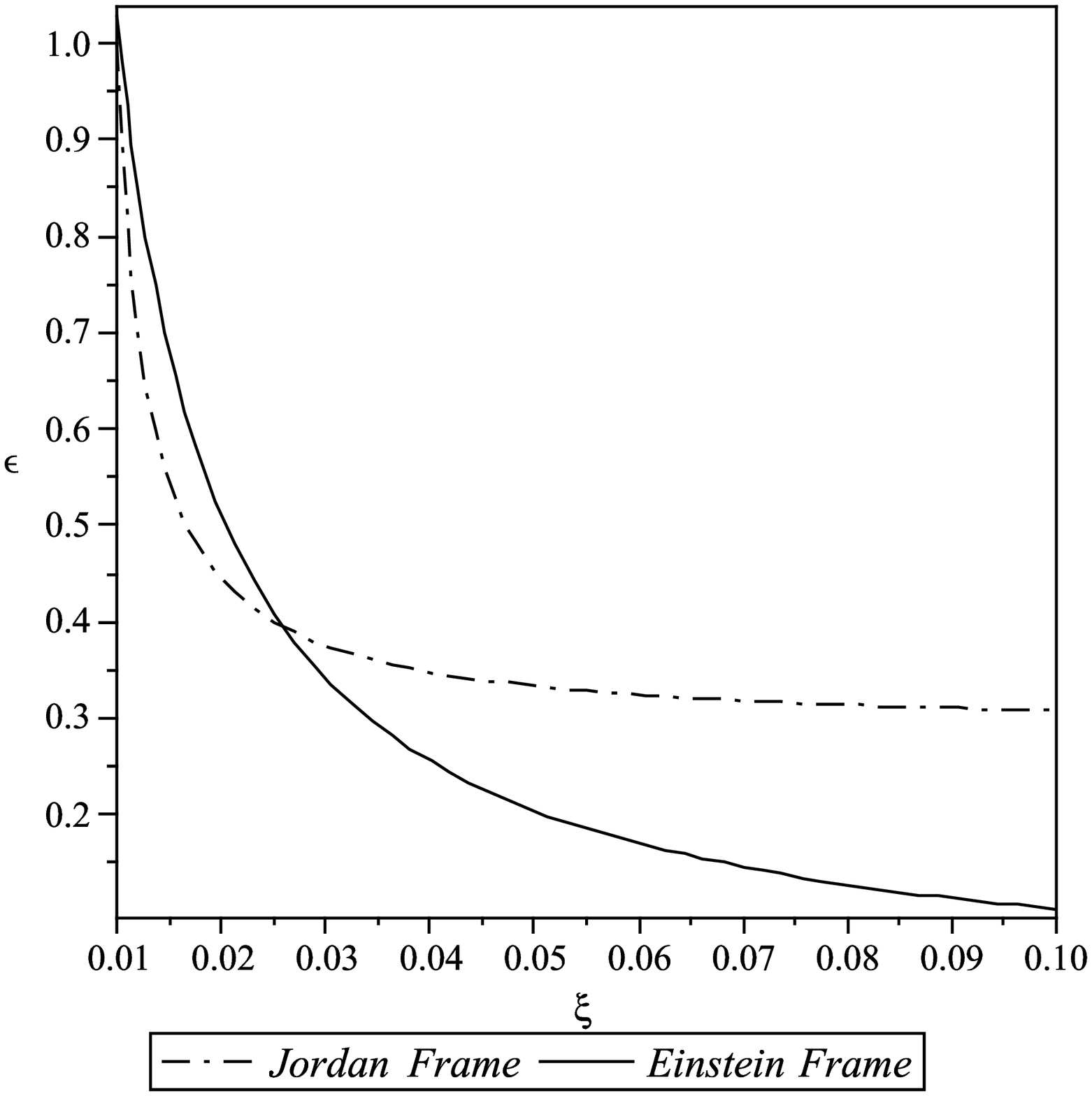}
\end{center}
\vspace{5 cm} \caption{\small { Comparison of two frames. In the
presence of the non-minimal coupling, natural exit from inflation
phase is possible without any additional mechanism. We have used
those values of $\xi$ which are supported by recent observations
[11]. }}
\end{figure}
Figure $2$ shows the variation of the spectral index versus $\xi$ in
the first and second order of the slow-roll parameters in the Jordan
frame. Generally, the second order results are more reliable than
the first order one in comparison with observation.
\begin{figure}[htp]
\begin{center}
\includegraphics{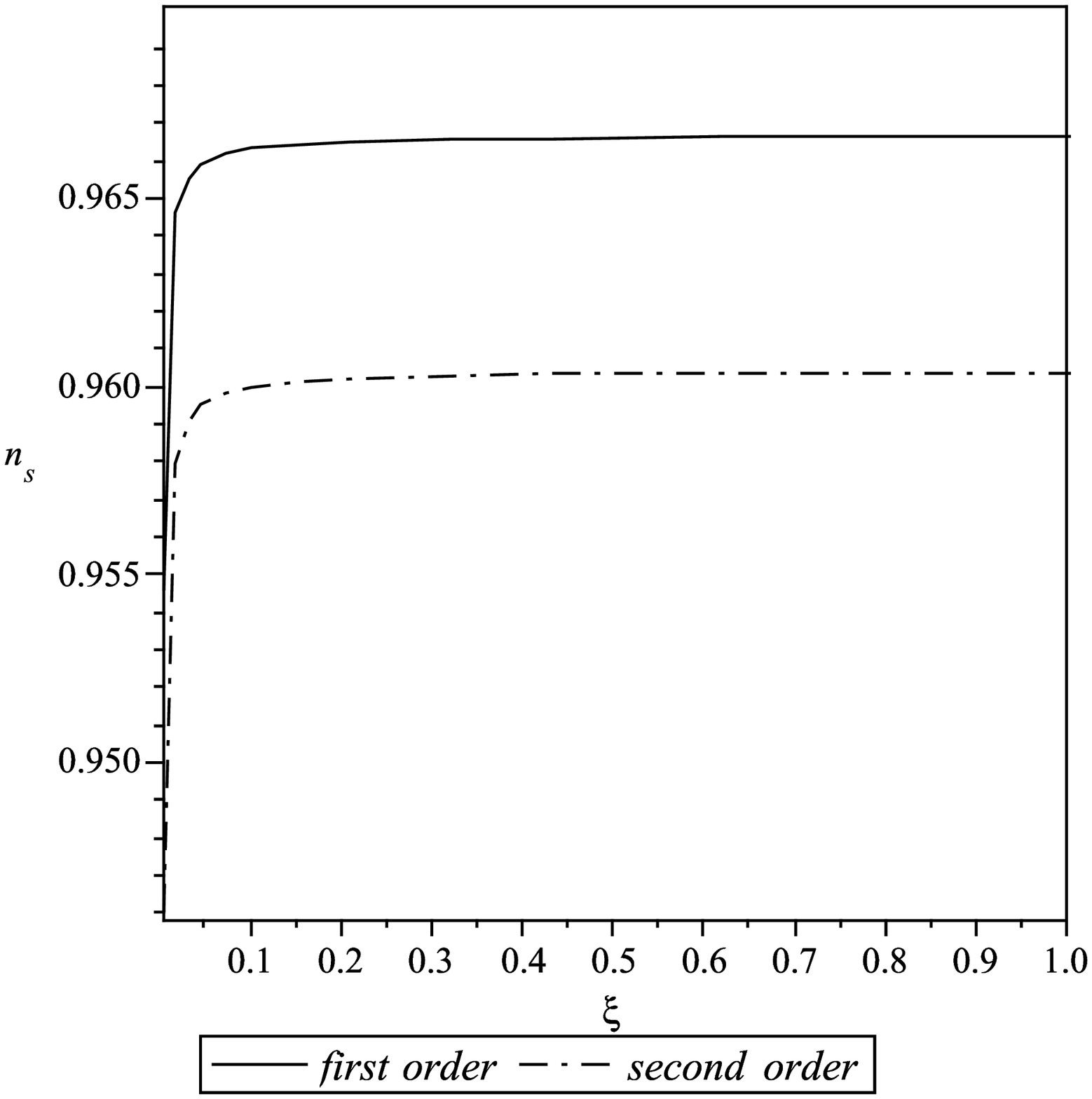}
\end{center}
\vspace{6cm} \caption{\small { Variation of the spectral index
versus $\xi$ in the first and second order of the slow-roll
parameters in the Jordan frame. }}
\end{figure}
Figure $3$ shows the variation of the spectral index versus $\xi$ in
the first and second order of the slow-roll parameters in the
Einstein frame. Once again, the second order results for large
values of $\xi$ are more reliable than the first order one in
comparison with observation.
\begin{figure}[htp]
\begin{center}
\includegraphics{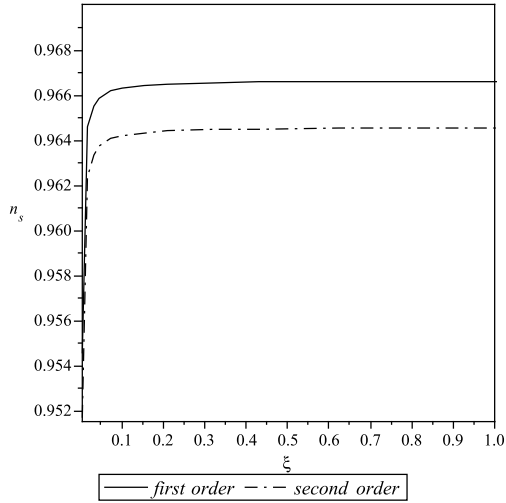}
\end{center}
\vspace{7 cm} \caption{\small {Variation of the spectral index
versus $\xi$ in the first and second order of the slow-roll
parameters in the Einstein frame. }}
\end{figure}
\newpage
Figure $4$ shows the variation of $n_{s}$ up to the second order of
the slow-roll parameters in two frames. As this figure shows, the
values of the $n_{s}$ in the Jordan frame for large values of $\xi$
are closer to observational data.
\begin{figure}[htp]
\begin{center}
\includegraphics{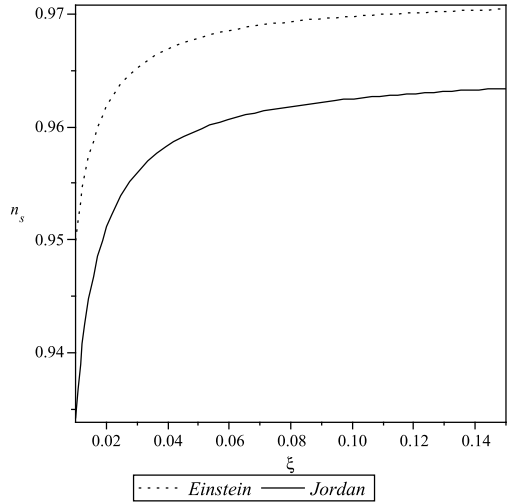}
\end{center}
\vspace{6 cm} \caption{\small { Variation of $n_{s}$ up to the
second order of the slow-roll parameters in two frames. }}
\end{figure}
Figure $5$ shows the running of the spectral index up to the second
order of the slow-roll parameters in two frames. As this figure
shows, the values of the $\alpha_{s}$ in the Jordan frame for large
values of $\xi$ are closer to observational data.
\begin{figure}[htp]
\begin{center}
\includegraphics{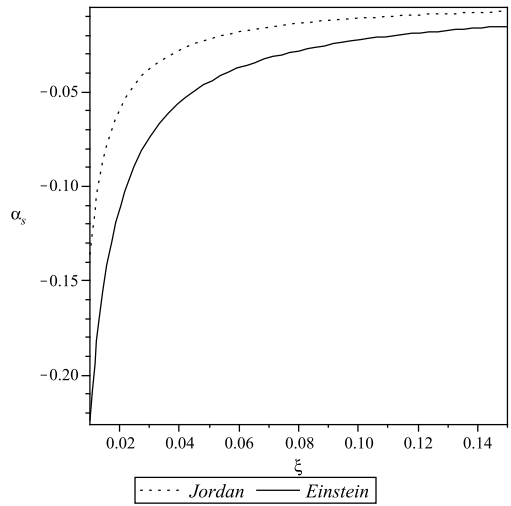}
\end{center}
\vspace{4 cm} \caption{\small { The running of the spectral index up
to the second order of the slow-roll parameters in two frames. }}
\end{figure}

\section{ Remarks on the Issue of Frames}
Although our analysis shows good agreements to the observational
data and automatic exit from the inflationary era in two frames, but
there are some differences in the results of calculations performed
in these two frames. Therefore, it is required to discuss about
possible fundamental physical meanings of each conformal frame and
the physical nature of differences. Since the two frames show
considerable differences between their results ( at least in the
second order calculations), they are not equivalent in some sense.
Then the important question arises: what frame is physically correct
for the description of the inflation? On the other hand, the Jordan
frame presents a variable effective gravitational "constant", while
the Einstein frame fixes the gravitational constant. The differences
related to the gravitational constant may lead to different
interpretations of the nature of the inertia and gravitation. In
which follows we address the above issues briefly: The Jordan and
Einstein frames are equivalent at the classical level, provided that
the units of mass, length, time, and quantities derived there from
scale with appropriate powers of the conformal factor in the
Einstein frame. What one measures is always the ratio of a quantity
to its units, which is the same in the two frames. In the Jordan
frame, the units are fixed and in the Einstein frame they vary with
the spacetime points. In the Jordan frame the gravitational
constant, $G$, is variable, in the Einstein frame it is fixed (but
now the units are changed)[13,14]. At the quantum level, the two
frames may be inequivalent. If we regard the conformal
transformation as a sort of canonical transformation, then
Hamiltonians that are classically equivalent (i.e., related by a
canonical transformation) are inequivalent when quantized: they have
different eigenfunctions and spectra of eigenvalues. So, it is
natural that the two frames are inequivalent at the quantum level
[14,15].

However, in this paper we are interested in the semiclassical
regime. We have computed the spectra of inflationary perturbations,
so there will be tensor and scalar fluctuations around a classical
background. In this semiclassical case, the two frames should be
inequivalent and the spectra (but not amplitudes) of inflationary
perturbations computed in the two frames turned out to be the same
in certain scenarios of inflation, to first order, but not to second
order ( see for instance [6]). As we have noticed in the literature
and also by private communications [15], it seems that it is easier
to compute fluctuations in the Einstein frame. There is also the
issue that, when we want to do a covariant quantization of the
Brans-Dicke-like scalar field, we can do it in the Einstein frame
but nobody has been able to do it in the Jordan frame [15].

In summary we can say that it seems that the Einstein frame is
preferred over the Jordan one at the semiclassical level studied
here. In some situations and to the first order, the two frames may
still be equivalent. Based on our calculations it seems that at the
second (and higher ) order, the two frames are inequivalent.
Nevertheless, the calculations in the Einstein frame are much
easier.

\section{Summary}
It is well-known that the results of calculations for inflation
quantities preformed in the Jordan and Einstein frames are the same
just to the first order of the slow-roll parameters. In the second
order calculations, there are considerable differences between
results of the calculations preformed in these two frames. In this
paper, we have studied a non-minimal inflation with calculations
performed up to the second order of the slow-roll parameters versus
the conformal coupling explicitly. Then we have compared our results
in two frames and also in two order of approximations. On the other
hand, by confronting these results with recent observations from
combined WMAP5+SDSS+BAO dataset, it is possible to see the
differences between two frames and two order of approximations in
this setup. Our analysis here shows that in the presence of the
non-minimal coupling, natural exit from inflation phase is possible
without adopting any additional mechanism. The second order results
have generally better agreement with WMAP5+SDSS+BAO ML result and
all calculated results lie in the range of the mean values. Our
analysis shows also that this non-minimal model generally gives
negative values of the running of the spectral index. Up to the
second order of the slow-roll parameters, the value of the running
of the spectral index in the Jordan frame is larger than the values
of the corresponding quantity in the Einstein frame. As our
numerical analysis shows, the second order results for large values
of $\xi$ are more reliable than the first order one in comparison
with observation. The values of the $n_{s}$ in the Jordan frame for
large values of $\xi$ are more acceptable in comparison with ML
observational data. From another view point we can use observational
constraints on the values of, for instance, $n_{s}$ to obtain a
constraint on the values of the non-minimal coupling $\xi$. Our
calculation up to the second order of the slow-roll parameters for
$n_{s}$ in the Jordan frame confronted with observation gives the
constraint $0.0003\leq\xi\leq 0.1666$. Nevertheless, we should
stress that for $V(\phi)\propto \phi^4$, viable cosmological models
exist even for
$\xi \gg 1$, see for instance [12].\\

In summary, in this paper the inflationary parameters of a
non-minimal inflation model are constrained by the observational
data and expressed in terms of the conformal coupling parameter,
which plays a fundamental role in the all of our calculations and
also suffers strong constraints from the observations on its
possible values. The analysis up to the second order of the
slow-roll parameters shows that there exist significative
differences between the results obtained in two conformally related
frames, and this is not the case for the first order approximations.
Our study shows that the results provided by the second order
calculations present better agreements to the observational data
than the first order. As an important result, we have  verified that
the non-minimally coupled inflationary model can provide automatic
exit from the inflationary period without the need to any extra
mechanisms.\\

{\bf Acknowledgement}\\

We would like to thank Professor Valerio Faraoni for his invaluable
comments on this work. In fact, the section $5$ of the present
manuscript is written based on a private communication with him.
Also we would like to thank an anonymous referee for his/her
critical comments on this work.

\end{document}